\documentclass[aps,prb,reprint,preprintnumbers,showpacs,twocolumn]{revtex4-2}

\usepackage{CJK}
\usepackage{longtable}
\usepackage{morefloats}
\usepackage[dvips]{graphicx}
\usepackage{color}
\usepackage{epsfig,graphicx,amsfonts,amsbsy}
\usepackage{amsmath,amsfonts,amsthm,amssymb}
\usepackage{appendix}
\usepackage{bbm}
\usepackage{makeidx}
\usepackage{url}
\usepackage{verbatim}
\usepackage[bookmarksnumbered,pdfpagelabels=true,plainpages=false,colorlinks=true,linkcolor=blue,citecolor=blue,urlcolor=blue]{hyperref}
\usepackage[rightcaption]{sidecap}
\usepackage{array}
\usepackage{booktabs}
\usepackage{multirow}
\usepackage{bbm}
\usepackage{floatrow}
\usepackage{tabularx}
\usepackage{cancel,soul,ulem}
\usepackage{subfigure}
\usepackage{graphicx}

\begin{document}
	\begin{CJK*}{UTF8}{gbsn}
		\title{Propagation, generation, and utilization of topologically trivial magnetic solitons in magnetic nanowires}
		\author{Kai-Tao Huang（黄铠涛）}
		\author{X.S. Wang}
		\email{justicewxs@hnu.edu.cn}
		\affiliation{School of Physics and Electronics, Hunan University, Changsha 410082, China}
		
		\begin{abstract}
			Magnetic solitons are nonlinear, local excitations in magnetic systems. In this study, we theoretically and numerically investigate the properties and generation of one-dimensional (1D) topologically trivial magnetic solitons in ferromagnetic nanowires. An approximate analytical soliton solution described by two free parameters is validated by comparing with the micromagnetic simulation. Across an interface between two media of different anisotropy, the reflection and refraction of a soliton are highly nonlinear that are different from the linear spin waves. A pair of magnetic solitons that propagate in opposite directions can be generated by alternately applying magnetic field or spin-polarized current pulses of opposite directions to at least two successive regions. Each soliton falls into a soliton solution that can be controlled by the generation process. These magnetic solitons can be used to drive domain wall motion over a certain distance determined by the soliton magnitude, allowing for discrete manipulation of domain walls compatible with the digital nature of information technology. Our findings pave the way for the application of topologically trivial solitons in spintronics.
			
		\end{abstract}
		\maketitle
	\end{CJK*}
	
	\section{Introduction}
	Over past few decades, spintronics devices have attracted much attention due to their low power consumption and non-volatile storage capabilities \cite{Lin2019,HIROHATA2020166711,Noel2020}. Some prototype devices such as racetrack memory \cite{Parkin2008,Zhang2015} and magnetic logic gate \cite{Cowburn2005,Luo2018,Luo2020} have been experimentally demonstrated. From both practical and academic perspectives, the properties and manipulation of magnetic solitons have garnered considerable interest.  Magnetic solitons are localized, nonlinear solutions of the magnetization dynamic equation that can maintain their shape and speed, in contrast to the linear wavepackets that suffer dispersion during propagation \cite{WP2024}. Magnetic solitons can be categorized based on the winding number of the magnetization in space, which serves as a topological invariant \cite{Wangbook}. Intensive studies have focused on topologically nontrivial solitons, such as domain walls, skyrmions, and hopfions, due to their exotic structures and potential applications \cite{Nagaosa2013,Thiaville2018,RF3,hopfion1,ferrimagnetdomain,circleskyrmion,vortexring}. The nontrivial topology of solitons usually implies enhanced stability of the soliton textures so that they are suitable for non-volatile data carriers. But it is also a double-edge sword that has some drawbacks, such as unexpected Hall effect \cite{skyrhall} and difficulty in creation \cite{DW2007,skyrmion2015}. 
    
    In contrast, topologically \textit{trivial} magnetic solitons are far less studied. Distinct from their topological counterparts, topologically trivial solitons in magnetic systems are solitary waves that are easy to be created and annihilated, referred to as magnetic drops \cite{KOSEVICH1990117}, magnetic droplets\cite{Jiang2024}, or just magnetic solitons if there is no ambiguity in the context \cite{Li2014}. Up to now, most studies are concerned about magnetic soliton in one-dimensional (1D) and two-dimensional (2D) systems. In 2D, the magnetic soliton can be created by using spin transfer torque underneath a nanocontact on a ferromagnetic film with perpendicular magnetic anisotropy (PMA) 	\cite{doi:10.1126/science.1230155,PhysRevLett.115.127205}. The 2D magnetic soliton is usually generated locally, and used as a nano-oscillator \cite{Macia2014}. The investigations on 1D magnetic soliton in ferromagnetic nanowires mainly focus on its analytical solutions and propagation \cite{PhysRevE.94.032222,PhysRevB.77.144416,LI2018390,LI2020166981,Zhao_2023,PhysRevB.109.014414}. However, there are still a lot of issues unsolved, mainly due to the complexity and nonlinearity of the Landau-Lifshitz-Gilbert (LLG) equation governing the magnetization dynamics. It is difficult to obtain exact analytical solutions for magnetic solitons, even in the simplest 1D case. So some approximations have to be made. The inhomogeneous demagnetization field is ignored, and dissipationless part of the LLG equation can be transformed into the nonlinear Schr{\"o}dinger 
	equation by keeping the cubic order of the soliton amplitude, so that analytical soliton solution solutions can be obtained \cite{PhysRevE.94.032222,PhysRevB.77.144416,LI2018390,LI2020166981,Zhao_2023,PhysRevB.109.014414}. But the validity of approximations above has not been fully justified, or in other word, whether the analytical solutions still work in the presence of intact demagnetization field and damping is not clear. How the solitons reflect at an interface between two nanowires with different material parameters is not known. Also, there is no systematical way to generate a certain soliton mode in a 1D nanowire yet.

	In this paper, we first verify the availability of the analytical soliton solution \cite{Li2014,PhysRevE.94.032222} in the presence of full demagnetization field, nonlinear effects and Gilbert damping by numerically simulation of the LLG equation starting from a certain solution. We find that for small soliton amplitude and low damping ($\sim 10^{-4}$), the analytical solution agrees well with the numerical results. When the soliton amplitude grows larger, the soliton speed becomes slower than the initial analytical solution. For a nanowire consist of two segments with different anisotropies, the reflection and refraction of an incident soliton is complicated and different from the linear spin waves. The soliton undergoes total refraction (reflection) when propagating into a medium with weak (strong) anisotropy, while refraction and reflection coexist at intermediate anisotropy. Then we design a method for generating magnetic solitons in a uniform domain by applying stimulation (magnetic field or spin polarization current pulses) to at least two successive regions. The directions of the stimulation in adjacent regions should be opposite. Two magnetic solitons are generated in pair and fall into two soliton solutions mentioned above. The properties of the generated solitons can be tuned by the strength and region width of the stimulation. When a magnetic soliton passes through a domain wall, the domain wall will shift a certain distance towards the opposite direction of the magnetic soliton propagation because of the angular momentum conservation \cite{PhysRevLett.107.177207}. The distance depends on the amplitude of the magnetic soliton. This procedure can be repeated to realize a discrete manipulation of domain wall, which is highly compatible with the digital nature of the information storage and transmission.
	
\section{Model and Results}
\subsection{Analytical solution of solitons and numerical verification}
\begin{figure}[!th]
		\centering
		\includegraphics[width=\textwidth]{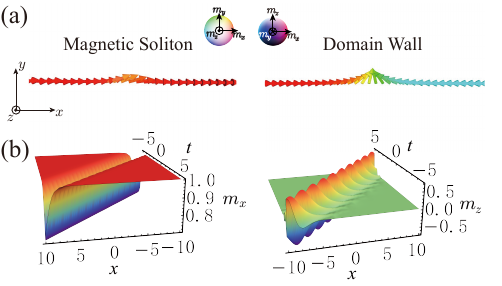}
		\caption{(a) Illustration of magnetization profiles of a typical magnetic soliton (left) and a head-to-head domain wall (right). The arrows denote the local magnetization direction. The color scheme is shown in the inset. (b) 3D plot of the analytical solutions of $m_x$ and $m_z$ components in $x-t$ plane.}
		\label{Fig1}
\end{figure}

A ferromagnetic nanowire of saturation magnetization $M_s$ along $x$ direction is considered. An easy-axis anisotropy along the wire maintains a single domain ground state of $m_x=\pm1$. We also allow an external field $\mathbf{B}_0=B_0 \hat{\mathbf{x}}$ to be applied along the wire. A magnetic soliton is a solitary excitation upon the single-domain ground state, in contrast to a domain wall which is the transition region between two opposite domains, as illustrated in Fig. \ref{Fig1}(a). The LLG equation reads,
\begin{equation}
	\frac{\partial \mathbf{m}}{\partial t}=-\gamma \mathbf{m} \times \mathbf{B}_\mathrm{eff}+\alpha \mathbf{m}\times \frac{\partial \mathbf{m}}{\partial t},  \label{LLG}
\end{equation}
where $\mathbf{m}$ is the unit vector of the magnetization field whose saturation magnitude is $M_s=0.86\times 10^6$ A/m, $\gamma=28$ GHz/T is the gyromagnetic ratio, and $\alpha$ is the Gilbert damping. $\mathbf{B}_\mathrm{eff}$ is the effective field that consists of the exchange field $\frac{2A}{M_s} \frac{\partial^2 \mathbf{m}}{\partial x^2}$ with the exchange constant $A$, the anisotropy field $\frac{2K}{M_s}m_x \hat{\mathbf{x}}$, the external field $B_0 \hat{\mathbf{x}}$, and the demagnetization field $\mathbf{B}_d$.
	
The analytical solution of the solitons is obtained under the following assumptions, (1) the demagnetization field $\mathbf{B}_d$ is approximated by shape anisotropy of infinite rod, $\mathbf{B}_d=\frac{1}{2}{\mu_0}M_s m_x \hat{\mathbf{x}}$, (2) the damping term can be ignored, and (3) the deviation of $\mathbf{m}$ from $x$ is small \cite{Li2014,PhysRevE.94.032222,PhysRevB.77.144416,LI2018390,LI2020166981,Zhao_2023,PhysRevB.109.014414}. To simplify the expressions, spatial unit $x_0=\sqrt{2A/\left(\mu_0 M_s^2\right)}$, temporal unit $t_0=(\gamma \mu_0 M_s)^{-1}$ and field unit $\mu_0M_s$ are used to define dimensionless variables and parameters $x'=(x-X_0)/x_0$, $t'=(t-T_0)/t_0$,  $\lambda=2K/\left(\mu_0 M_s^2\right)$, and $b_0=B_0/(\mu_0M_s)$. Here $X_0$ and $T_0$ are certain initial position and time, respectively. By defining $\psi=m_y-i m_z$ and keeping the cubic terms of $\psi$, we obtain a Schr\"{o}dinger equation with both linear and nonlinear terms, 
\begin{equation}
		{i \frac{\partial \psi}{\partial t'} =- \frac{\partial^2 \psi}{\partial x'^2}+\left(b_0+\lambda+\frac{1}{2}\right)\psi - \frac{1}{2}\left(\lambda+\frac{1}{2}\right)|\psi|^2\psi=0.} \label{NLS}
\end{equation}
When $b_0+\lambda+\frac{1}{2}=0$, the rest is a pure nonlinear Schr\"{o}dinger equation and one of its soliton solutions has been obtained by the Hirota method \cite{Li2014,LI2018390,MA2022100220}. Otherwise, the linear term provides an extra precessing frequency around $x$ direction. Except arbitrary initial position $X_0$ and time $T_0$, the exact solutions of Eq. \eqref{NLS} are determined by only two free parameters $a$, $b$ as
	\begin{equation}
		\psi(x',t')=
		C(a)e^{i[bx'-\omega_0(a,b)t']}
		\mathrm{sech}[a(x'-2bt')-\ln2C(a)],
		\label{psi_solution}
	\end{equation}
where $C(a)=\frac{2a}{\sqrt{\lambda+1/2}}$ and $\omega_0=h_0+\lambda+\frac{1}{2}-(a^2-b^2)$. Physically, this solution has an oscillatory part $e^{i[bx'-\omega_0(a,b)t']}$ with a precessing frequency $\omega_0/t_0$ and ``wavelength" $2\pi x_0/b$, and an envelope $C(a)\mathrm{sech}[a(x'-2bt')-\ln2C(a)]$ with amplitude $C(a)$ and propagating speed $2bv_0$ ($v_0=x_0/t_0$ is the unit of speed). $a$ is also inversely proportional to the width of the envelope. This solution is a propagating one itself without any driving force, since the system is energy conservative.

The three components of $\mathbf{m}$ are derived from $\psi$ as $m_x=\sqrt{1-|\psi|^2}\equiv\psi_x$, $m_y=\mathrm{Re}\psi\equiv\psi_y$ and $m_z=-\mathrm{Im}\psi\equiv\psi_z$. The profiles of $m_x$ and $m_z$ are plotted in Fig. \ref{Fig1}(b) for $a=b=1$. From Eq. \eqref{psi_solution} and Fig. \ref{Fig1}(b), $m_x$ behaves as a dark soliton with amplitude $1-\sqrt{1-C^2}$ without oscillation and $m_{y,z}$ behave as bright solitons precessing with circular frequency $\omega_0/t_0$. 
	
\begin{figure}[!ht]
		\centering
		\includegraphics[width=\textwidth]{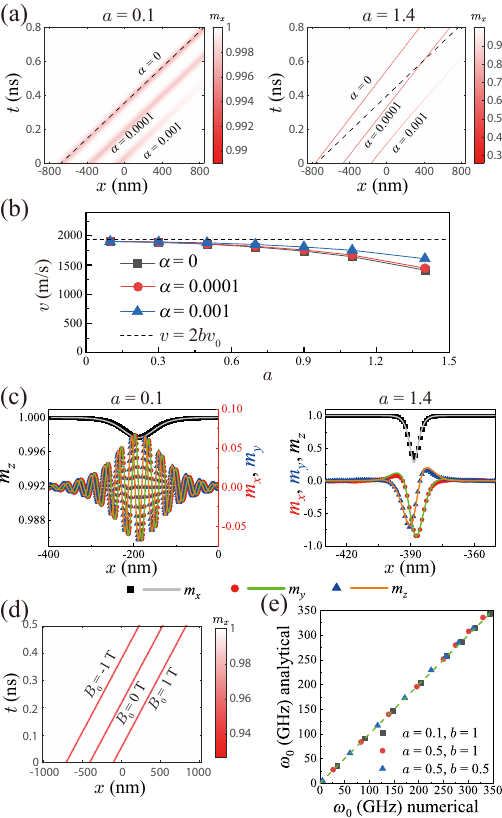}
		\caption{(a) Density plots of $m_x$ in $x-t$ plane for $b=1$ and different $a=0.1$ (left) and 1.4 (right). (b) Soliton speed for different $a$ and damping $\alpha$. Symbols are numerical results. The dashed line is the theoretical value $v=2bv_0$.
			(c) Magnetization profiles at $0.2$ ns for $b=1$ and different $a=0.1$ (left) and 1.4 (right). Symbols are the numerical data and solid lines are the analytical formula.
            (d) Density plots of $m_x$ for different external field $H$.
            (e) Comparison of the precession frequency for different $a$ and $b$. The horizontal and vertical axes are numerical and analytical results, respectively. The green dashed line is $y=x$.}
		\label{Fig2}
\end{figure}

The analytical solution Eq. \eqref{psi_solution} is a precise solution of the nonlinear Schr\"{o}dinger equation \eqref{NLS}, which is valid only if the three assumptions above can be justified. Thus, we hope to know to what extent the solution is valid in the LLG dynamics, with the intact demagnetization field, nonlinear terms, and damping. To do this, the analytical solution Eq. \eqref{psi_solution} at $t=0$,
\begin{equation}
		\psi(x',0)=
		C(a)e^{ibx'}
		\mathrm{sech}[ax'-\ln2C(a)],
		\label{initial}
	\end{equation}
is set to be the initial state, and Mumax3 package \cite{mumax3} is used to simulate the LLG equation \eqref{LLG} to see the time evolution of the initial state and find out the difference from the expected solution \eqref{psi_solution}. The parameters are chosen to be $B_0=0$, $A=12$ pJ/m, and $\lambda=8$. The nanowire is $2048$ $\rm{nm}$ $\times 1$ $\rm{nm}$ $\times$ $1$ $\rm{nm}$ and the mesh size is $1$ $\rm{nm}$ $\times$ $1$ $\rm{nm}$ $\times$ $1$ $\rm{nm}$, which is small enough compared to the exchange length scale $x_0\approx 5.08$ nm.  Figure \ref{Fig2}(a) shows the density plots of $m_x$ component for fixed $b=1$, $a=0.1$ (left) and $1.4$ (right) in the $x-t$ plane for $\alpha=0$, 0.0001, and 0.001. For brevity and better comparison, we artificially stack the results for different $\alpha$ in the same subfigures. All the solitons move in a same constant speed in each subfigure. The soliton amplitude becomes larger and its width becomes thinner when $\alpha$ increases.  For $\alpha=0.001$, the soliton amplitude obviously decays for each case. However, for $\alpha=0.0001$, within the micrometer length scale and the nanosecond time scale, the decay is unobvious. Since the damping of magnetic insulators such as YIG can be as low as $10^{-5}$ \cite{YIGdamping}, the decay of the soliton is acceptable in a length scale of $\mu$m. The dashed line in each subfigure is the analytical formula $x=2bv_0 t$, where the speed $2bv_0=1932$ m/s is quite fast in nano-devices. We can see for $a=0.1$, the simulated speed is almost the same as the analytical solution. But for $a=1.4$, the resultant speed is slower than the analytical one. This is understandable because for $a=1.4$, the amplitude $C=0.96$ is already very large, so the ignorance of higher-order nonlinearity leads to more significant deviation from the analytical solution. We further compare the simulated speed $v$ for different $a$ and different $\alpha$ in Fig. \ref{Fig2}(b) for fixed $b=1$. For small amplitude, the deviation from the analytical value is small, while the deviation increases with $a$. Also, larger damping slightly accelerates the motion of the soliton. We then compare the details of the magnetization profile of the solitons.
In Fig. \ref{Fig2}(c), numerical results of the three magnetization components at 0.2 ns are plotted in symbols. Solid lines are analytical solutions. Here, because the speed is not precisely $2bv_0$, a properly shifted $X_0$ is used to solely compare the profile. For $a=0.1$, the agreement between the numerical and analytical profile is very good. For $a=1.4$, the agreement is still good, but the discrepancy is more apparent than that in the case of $a=0.1$.  
	
According to the analytical solution, the longitudinal magnetic field $B_0$ does not affect the soliton shape and speed, but changes the precession frequency $\omega_0$. In Fig. \ref{Fig2}(d), we change different $B_0$ ($B_0 =-1$ T, 0, 1 T) and simulate the evolution from the initial state of $a=0.5$, $b=1$, and stack the numerical data in the same figure. The three solutions are almost the same in $m_x$ component. We further compare the precession frequency of $m_{y,z}$. In Fig. \ref{Fig2}(e), the horizontal axis records the precession frequency from the Fourier transform of the numerical data,  and the vertical axis is the analytical formula $\omega_0 =b_0+\lambda+\frac{1}{2}-(a^2-b^2)$. The agreement is very good.

To summarize this subsection, the analytical solution Eq. \eqref{psi_solution}  of the nonlinear Schr\"{o}dinger equation (as an approximation of the LLG equation) is compared with the full numerical simulation results. The deviation from the analytical solution, manifested mainly in the slower propagation speed, becomes larger when the soliton amplitude increases, which may originate from the ignorance of the higher-order terms. The soliton profile and the precession frequency still agree well with the analytical solution. In the following sections, we let $B_0=0$ and $\alpha=0$ without losing generality.

\subsection{Reflection and refraction of magnetic solitons at an interface}\label{interface}

We now consider the soliton propagation in a 1D nanowire heterostructure
which may be useful in devices. There are quite a few studies on the reflection and refraction of linear spin waves, which follow a generalized Snell's law like that in linear optics \cite{snell1,snell2,Xiaojiang}. However, for nonlinear solitons, the behaviors may be very different. As an analogy, optical solitons, which can be described by nonlinear Schr{\"o}dinger equation \cite{optic1,optica2} or nonlinear Helmholtz equation \cite{optica3,Hosoliton}, have been intensively investigated in nonlinear optics \cite{JDGibbon_1972,PJCaudrey_1973,Boardman1997,Andrei,LIN2016}. The reflection and refraction of optical solitons are shown to be much more complicated than that in linear media \cite{Sanchez-Curto:07,Sanchez-Curto:10,opticasnell}. Therefore, it is essential to investigate the reflection and refraction behaviors of magnetic solitons.

\begin{figure}[!ht]
	\centering
	\includegraphics[width=\textwidth]{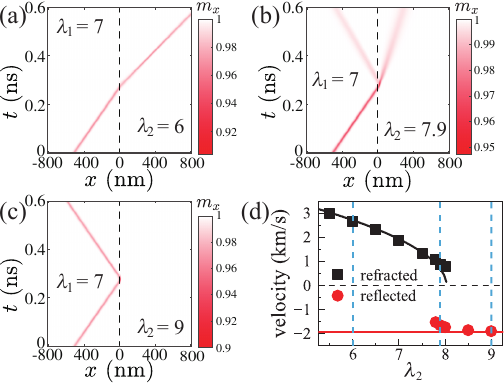}
	\caption{Behaviors of a soliton of $a=0.3$, $b=1$ propagating towards an interface at $x=0$ from the left half with $\lambda_1=7$. (a)-(c) Density plots of $m_x$ for anisotropy of the right half to be (a) $\lambda_2=6$, (b) $\lambda_2=7.9$, and (c) $\lambda_2=9$. (d) Dependence of refracted (black squares) and reflected (red dots) soliton speed for different anisotropies $\lambda$ of the right half. The solid lines are the analytical formula. The vertical dashed lines indicate $\lambda_2=6$, 7.9, and 9.}
	\label{Fig3}
\end{figure}

1D nanowire heterostructures of different materials have been experimentally realized \cite{nanowirereview}. For simplicity, we consider an interface at $x=0$, between two segments with different anisotropies that can be tuned by electric field \cite{anisotropy2,anisotropy} or doping \cite{anisotropy1}. The anisotropy is $\lambda_{1}$ in the left half and $\lambda_2$ at the right half, while the other parameters are the same as above subsection. The incident soliton is prepared by setting the initial state Eq. \eqref{initial} with $a=0.3$, $b=1$. Figure \ref{Fig3} (a)-(c) show the density plots of $m_x$ of the incident $a=0.3$, $b=1$ soliton for fixed $\lambda_1=7$ and $\lambda_2=6$, 7.9 and 9, respectively. We see that for $\lambda_2=6$, there is almost no reflection, and the soliton goes into the right half with a slightly larger speed. For $\lambda_2=9$, on the contrary, a total reflection is observed. The refracted and reflected solitons coexist when $\lambda_2=7.9$. Figure \ref{Fig3}(d) summarizes the refracted (black squares) soliton speed and the reflected (red dots) soliton speed for different $\lambda_2$. There is a narrow region that clear refracted soliton and reflected soliton can be both identified. Clearly, the behavior is distinct from the linear spin waves \cite{snell1,snell2,Xiaojiang} (see Appendix B).

For the total refraction regime at small $\lambda_2$, we suppose that the incident $a=0.3$, $b=1$ soliton turns into another soliton solution with $a_2$, $b_2$ after passing the interface without loss. Then, since the nanowire is rotationally symmetric around $x$ axis, the $x$ component of the total angular momentum $L_x$ should be conserved. For a soliton with solution $\psi$, the $x$ component of angular momentum of a soliton deviated from the uniform domain is
\begin{equation}
	L_x=\gamma^{-1}M_s S \int_{-\infty}^{\infty}(1-\psi_x) dx,
\end{equation}
where $S$ is the cross-section area of the nanowire. For small amplitude, we employ the approximation $\psi_x=\sqrt{1-|\psi|^2}\approx 1-|\psi|^2/2$, the integral can be carried out analytically, 
\begin{equation}
    \int_{-\infty}^{\infty}(1-\psi_x) dx\approx \int_{-\infty}^{\infty}\frac{|\psi|^2}{2} dx =\frac{4a x_0}{\lambda+1/2},
\end{equation}
so the conservation gives
\begin{equation}
    \frac{a}{\lambda_1+1/2}=\frac{a_2}{\lambda_2+1/2}. \label{arela}
\end{equation}
On the other hand, in this regime, we find from the numerical data that the precession frequency $\omega_0$ is almost the same at both sides of the interface, which is natural for linear waves but not necessarily true for nonlinear solitons. Thus, we have the second relation,
\begin{equation}
    \lambda_1-(a^2-b^2)=\lambda_2-(a_2^2-b_2^2). \label{brela}
\end{equation}
Solving for $a_2$ from Eq. \eqref{arela} $a_2$ and combining Eq. \eqref{brela}, $b_2$ can be solved as
\begin{equation}
b_2=\sqrt{\lambda_1-\lambda_2+a^2\left[\left(\frac{\lambda_2+1/2}{\lambda_1+1/2}\right)+1\right]+b^2}. \label{bana}
\end{equation}
The refracted soliton speed is $2b_2v_0$, plotted in Fig. \ref{Fig3}(d) by the black solid line, which shows good agreement with the numerical data. 
In the total reflection regime at large $\lambda_2$, the reflected soliton should have $a_1=a$, $b_1=-b$. The red solid line in Fig. \ref{Fig3}(d) labels
$v=-2bv_0$. The agreement is good for large $\lambda_2$.

The radicand in \eqref{bana} is non-negative, giving a critical value of $\lambda_2$ for total refraction, which is $\sim8.02$ here. But indeed, near the critical value, the behavior becomes highly complex. This complexity arises not only from the coexistence of reflection and refraction, but also from the conversion into linear spin waves (as illustrated in Appendix A).

We can intuitively understand the above behaviors as follows. When an incident soliton from one region hits an interface, since it is no longer the solution for the other region, the system has to find a new solution, so reflection and refraction occur. If the anisotropy of the other section is weaker (i.e. a softer magnet), the spins therein are easy to be excited, so the soliton tends to totally enter the softer region, giving a total refraction Fig. \ref{Fig3}(a). On the other hand, for a very large anisotropy (i.e. a hard magnet), it is difficult to excite the spin precession, so the soliton cannot enter the harder region, giving a total reflection Fig. \ref{Fig3}(c). In the intermediate region, the case becomes very complicated, not only because of the coexistence of reflected and refracted solitons, but also because of the excited linear spin waves. In this process, the energy, angular momentum and precession frequency are no longer conserved for the solitons because of the existence of linear spin waves. Further investigations are needed to fully understand this behavior, and will be the subject of future studies.

	\subsection{Generation of magnetic solitons}

	Practically, it is more important to generate a soliton in an efficient and simple way. A quite intuitive idea is to use a magnetic field pulse or a spin current pulse to locally stimulate the magnetization to tilt. However, after a plethora of tests, we find that applying a stimulation in a single region cannot generate a propagating soliton. Depending on strength, direction and duration of the stimulation, there can be irregular excitations dispersing into linear spin waves, nucleation of a new domain, or locally precessing soliton (i.e. soliton of $b=0$). Only when stimulation of different directions is applied in at least two regions, propagating solitons can be generated in pairs.

\begin{figure}[!ht]
		\centering
		\includegraphics[width=\textwidth]{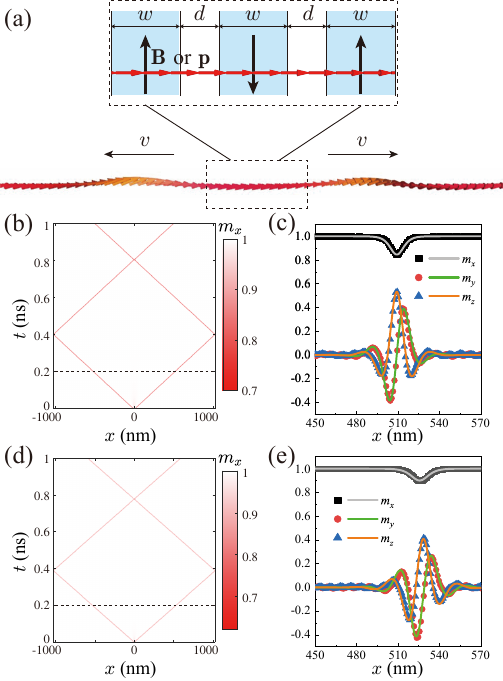}
		\caption{(a) Schematic diagram of the soliton generation apparatus. (b)(d) Density plots of $m_x$ in the $x-t$ plane with a pair of solitons generated at the center of the wire. The dashed lines indicate 0.2 ns.
			(c)(e) The magnetization profiles of the rightward solitons at 0.2 ns. The symbols are the numerical data and the solid lines are the analytical formula.  (b)(c) magnetic field stimulation; (d)(e) spin current stimulation.}
		\label{Fig4}
	\end{figure}

	Figure. \ref{Fig4}(a) shows the set-up of the apparatus. We apply magnetic field $\mathbf{B}$ perpendicular to the wire in at least two successive regions of width $w$ (here, we use three regions as an example). The directions of the field in adjacent regions are opposite. The magnetic field can also be  replaced by spin-polarized current perpendicularly injected into the wire. The spin polarization directions should also be opposite in adjacent regions. There can be a gap of width $d$ between adjacent regions considering the convenience of production in reality.

	We first consider the case that the amplitude of the stimulation (magnetic field or spin current) in each region is the same. In this case, the whole range of stimulation is mirror-symmetric for regions of odd number, and inversion-symmetric for regions of even number. After applying a field pulse of large magnitude (several Tesla) but short time ($\sim$picoseconds), a pair of solitons can be generated, and they propagate towards opposite directions with the same speed. A typical numerical result is exhibited by density plot of $m_x$ in Fig. \ref{Fig4}(b). The three regions of width $w=10$ nm are put at the center of the wire, and the separation $d$ is 0. The material parameters are $A=13$ pJ/m, and $\lambda=8$. The magnitude of the field pulse is $B=4$ T and the duration is 1 ps. These parameters will be used throughout this section unless otherwise specified. It is clear that a pair of solitons is generated. The two solitons propagate in opposite directions at the same constant speed. The solitons are reflected by the boundaries and pass through one another maintaining the same shape, amplitude, and speed, which are typical behaviors of solitons. At 0.2 ns [the dashed line in Fig. \ref{Fig4}(b)], we plot the magnetization profile of the right hand side in component form (shown by symbols). The solid lines are the analytical solutions of $a=0.777$ and $b=1.262$, with small initial position and time shift $X_0=-0.47$ nm and $T_0=0.62$ ps, obtained by fitting the numerical data with the analytical formula. The fitted analytical profile agrees very well with the numerical data. The speed also agrees well with $v=2bv_0=2538$ m/s, which is quite fast in ferromagnetic systems. Similarly, the soliton at the left hand side also agrees well with the solution of $b=-1.262$, giving the same speed and negative moving direction. 
	If we use spin-polarized current injected in $z$ direction of density $1.7\times 10^{13}$ A/m$^2$ and polarization $P=0.5669$ instead of the magnetic field, the results are plotted in Fig. \ref{Fig4}(d) and (e). A pair of solitons with slightly larger speed and smaller amplitude ($a=0.635$, $b=1.317$ from fitting) is stimulated. In this case, the magnetization is excited via the Slonczewski spin transfer torque \cite{STT1,STT2}.

	\begin{figure}[!ht]
		\centering
		\includegraphics[width=\textwidth]{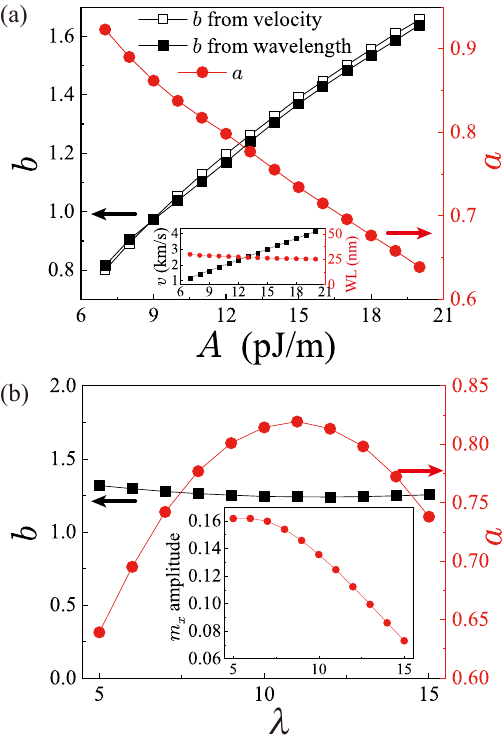}
		\caption{(a) Dependence of $b$ (left axis) and $a$ (right axis) on the exchange $A$. The solid square and the hollow square are obtained from the speed formula $v=2b v_0$ and the wavelength $2\pi x_0/b$, respectively. Inset: dependence of the soliton speed (left axis) and the wavelength (right axis) on $A$. (b) Dependence of $b$ (left axis) and $a$ (right axis) on the anisotropy $\lambda$. Inset: dependence of $m_x$ amplitude of the soliton on $\lambda$. }
		\label{Fig5}
	\end{figure}
	
	We further investigate how to control the generated solitons. We consider the influence of material parameters (exchange constant $A$ and anisotropy $\lambda$), as well as the stimulation conditions (stimulating field strength $B$ and width of stimulating region $w$). We choose $A=13$ pJ/m, $\lambda=8$, $B=4$ T and $w=10$ nm as the reference. We alter each parameter with other parameters fixed. $d$ is fixed to 0 and pulse duration is fixed to 1 ps except otherwise specified. 
    
    Firstly, we discuss the dependence of $a$ and $b$ on material parameters $A$ and $\lambda$.  Figure \ref{Fig5}(a) shows the dependence of $a$ (right axis) and $b$ (left axis) on the exchange constant $A$. Since stronger exchange interaction leads to tendency of more collinear magnetization alignment, the stimulated soliton amplitude as well as $a$ decreases with increasing $A$.
	On the other hand, $b$ increases with $A$ so the stimulated solitons propagate faster when $A$ is larger. In the inset of Fig. \ref{Fig5}(a), we plot the speed and the wavelength of the soliton against $A$. The wavelength $2\pi x_0/b$ (right axis) does not change much with $A$, so $b$ is approximately proportional to $x_0\propto \sqrt{A}$. Thus, the speed $v=2bv_0\propto b\sqrt{A}$ is almost linear to $A$, which agrees well with the numerical result of speed (left axis). Figure \ref{Fig5}(b) shows how the anisotropy $\lambda$ affects the stimulated soliton. The left axis shows that $b$ (as well as the wavelength and the speed) is insensitive to $\lambda$. $a$ is not monotonic with respect to $\lambda$, but the soliton amplitude decreases with increasing $\lambda$. This is because larger anisotropy also prefers collinear magnetization alignment along $x$, and the $m_x$ amplitude $1-\sqrt{1-C^2}$ depends on both $a$ and $\lambda$. 
	
	\begin{figure}[!ht]
		\centering
		\includegraphics[width=\textwidth]{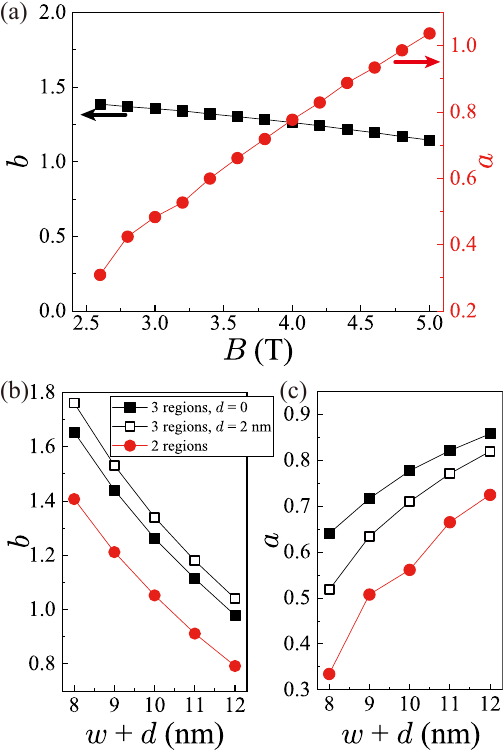}
		\caption{(a) Dependence of $b$ (left axis) and $a$ (right axis) on the stimulation field strength $B$. (b)(c) Dependence of $b$ (b) and $a$ (c) on the the stimulation width plus the region separation $w+d$. Black solid squares, black hollow squares, and red dots are 3-region stimulation with $d=0$, 3-region stimulation with $d=2$ nm, and 2-region stimulation with $d=0$, respectively.}
		\label{Fig6}
	\end{figure}

	We then discuss the influence of the stimulation conditions on the soliton properties. Figure \ref{Fig6}(a) shows how $b$ (left axis) and $a$ (right axis) depend on the stimulation field strength $B$. It is clear that $a$ is positively related to the stimulation strength in a nearly linear manner. $b$ does not depend on the stimulating field strength $B$ very much. It only slightly decreases when $B$ increases. The possible reason will be discussed later.
	When the region width $w$ is altered, $b$ as well as the wavelength $2\pi x_0/b$ depends on $w$, as shown in Fig. \ref{Fig6}(b) by the black solid squares. $b$ decreases with the region width, so that the wavelength $2\pi x_0/b$ increases with the region width. We also repeat the calculation for a finite separation $d=2$ nm,  and for the stimulation using two regions keeping $d=0$. We find that for same $w+d$, a finite separation does not change $b$ very much (less than $10\%$ larger). But the two-region stimulation gives smaller $b$, so that the wavelength is larger and the speed is smaller. The change of $a$ with region width is shown in Fig. \ref{Fig6}(c). Larger region excites soliton of larger amplitude. Furthermore, a finite gap slightly reduces the stimulated soliton amplitude, while two-region stimulation can only induce much weaker solitons. That is why we use three-region stimulation as the example in this section. 
	
	We try to have some physical insights into the stimulation process according to the numerical results. From the analytical solution we can see there is a traveling wave factor $e^{i[bx'-\omega_0(a,b)t']}$, so within the range of the soliton, $m_y$ and $m_z$ are oscillatory. The alternating field pulses force the magnetization to form a locally oscillating profile with wavelength larger than $2w$. Thus, the wavelength of the soliton is mainly related to the width of stimulation region [Fig. \ref{Fig6}(b)], and does not depend much on the material parameters like $A$ and $K$, as we find numerically in Fig. \ref{Fig5}. Because the field directions are opposite in adjacent regions, larger region width $w$ helps to release the strong exchange energy near the region boundaries, so the soliton amplitude can be larger. $w$ cannot be too small. If $w$ is closed to the exchange length $x_0$, it is hard to effectively drive the magnetization to tilt. On the other hand, the soliton amplitude depends not only on the strength of the stimulation field but also on the material parameters $A$ and $K$, which is natural according to the LLG equation. 
	
	To be more practical, we perform further numerical tests with respect to realistic conditions. We replace the square field pulse of width 1 ps with a gaussian field pulse of full width at half maximum 1 ps. All the results remain qualitatively unchanged. We also try the stimulation of solitons in magnetic strips without rotation symmetry around the $x$ axis. For a thin strip not much wider than the exchange length $x_0$, soliton pairs can still be stimulated, although the analytical solution Eq. \eqref{psi_solution} is no longer valid. Note that along with the solitons, other ``byproducts" are inevitably excited at the same time. For example, in Fig. \ref{Fig4}(b) and (d), there are light red regions near the source. Also, in Fig. \ref{Fig4}(c) and (d), there are still oscillations outside the range of solitons. Indeed, the unwanted excitations dissipate into linear spin waves and spread out. For similar soliton amplitude, two-region stimulation with stronger field or current can significantly reduce the unwanted excitations.

	\begin{figure}[!ht]
		\centering
		\includegraphics[width=\textwidth]{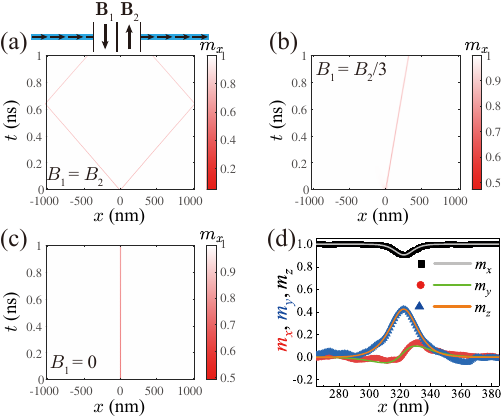}
		\caption{(a) Set-up of two region stimulation and the density plot of $m_x$ for $B_1=B_2=6$ T. (b)(c) Density plots of $m_x$ for (b) $B_1=B_2/3$ and (c) $B_1=0$. (d) Magnetization profile at 1 ns for $B_1=B_2/3$. Symbols are the numerical data and solid lines are the fit by the analytical solution Eq. \eqref{psi_solution}.}
		\label{Fig7}
	\end{figure}
	
	Now we allow the amplitudes of the magnetic fields to be different, so that the stimulation is asymmetric. We consider two regions shown in Fig. \ref{Fig7}(a) for clarity, in which $\mathbf{B}_1$ ($\mathbf{B}_2$) in the left (right) region points to $-y$ ($+y$). We keep $A=13$ pJ/m, $\lambda=8$, $w=10$ nm, $d=0$, pulse duration 1 ps and $B_2=6$ T. When $B_1=B_2$, symmetric soliton pairs are generated, as we discussed above, shown in Fig. \ref{Fig7}(a). As $B_1$ decreases, the soliton pairs become asymmetric and the speed becomes slower. The amplitude of the leftward soliton becomes smaller. For $B_1=B_2/3=2$ T, there is only one observable soliton propagating rightward at a much slower speed, as shown in Fig. \ref{Fig7}(b). Indeed, there is a rightward soliton of very small amplitude, along with other noise excitations. Figure \ref{Fig7}(d) exhibits the magnetization profile at 1 ns. The symbols are numerical data and the solid lines are the fit using the analytical solution Eq. \eqref{psi_solution} with $a=0.6221$ and $b=0.1596$. It is clear that there is much more significant noise that the symmetric stimulation results in Fig. \ref{Fig4}(c). For $B_1=0$, only one region is stimulated and there is no propagating soliton, as shown in Fig. \ref{Fig7}(c).

	\subsection{Discrete manipulation of domain wall by solitons}
		\begin{figure}[!ht]
		\centering
		\includegraphics[width=\textwidth]{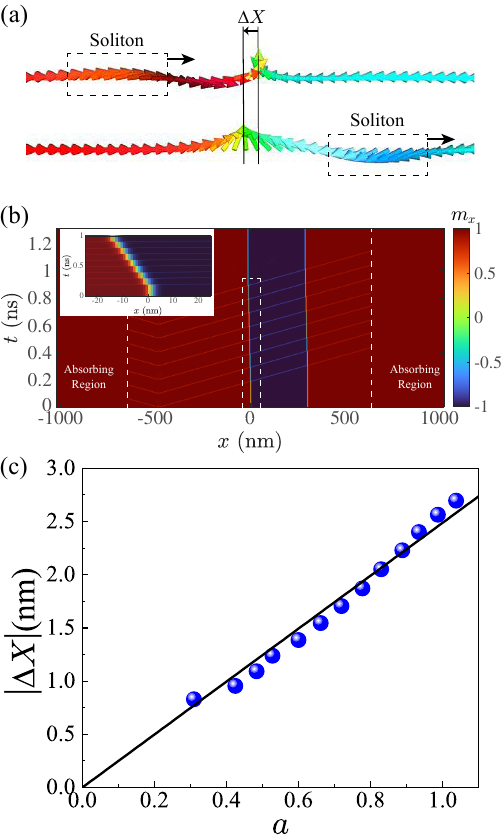}
		\caption{(a) Schematic diagrams for a soliton passing through a domain wall. (b) Density plot of $m_x$ in $x-t$ plane. Two domain walls are initially placed at the center and $x=300$ nm. Solitons are stimulated at $-500\sim-470$ nm with a period of 0.101 ns. Inset: close-up near the domain wall. (c) Comparison between the domain wall shift obtained from the numerical data (balls) and the analytical formula $|\Delta X|=4ax_0/\left(\lambda+\frac{1}{2}\right)$ (solid line).}
		\label{Fig8}
	\end{figure}
It has been well known that the linear spin wave can drive domain wall propagation \cite{PhysRevLett.107.177207,science1121,science1125}. For a uniaxial magnetic wire, the rotational symmetry around $x$ direction maintains the conservation of the $x$-component of the angular momentum. Similar to the linear spin wave \cite{PhysRevLett.107.177207}, the soliton also carries finite angular momentum. Thus, a magnetic soliton passing through a domain wall can also drive the domain wall to move to the opposite direction of the soliton propagation. Because of the locality of the soliton, the domain wall has a finite displacement of $\Delta X$ after the penetration of one soliton, as illustrated in Fig. \ref{Fig8}(a) and in the animation in Supplementary Material \cite{SM}. We consider a soliton described by Eq. \eqref{psi_solution} in a wire of length $L$, and the $m_x$ profile of the domain wall is $f_\text{DW}(x-X_\text{DW})$, where $X_\text{DW}$ is the position of the domain wall center and $X_\text{DW}=\frac{1}{2}\int_{-L/2}^{L/2}f_\text{DW}(x-X_\text{DW})dx$ \cite{Donahue2004}.  From the angular momentum conservation before and after the soliton passes through the domain wall $\int_{-L/2}^{L/2}m_x dx=\text{constant}$, we assume that the soliton and the domain wall are far from each other so that the integration is can be separated, 
	\begin{multline}
	    \int_{-L/2}^{L/2}f_\text{DW}(x-X_1)dx-\int_{-L/2}^{L/2}(1-\psi_x)dx\\
        =\int_{-L/2}^{L/2}f_\text{DW}(x-X_2)dx+\int_{-L/2}^{L/2}(1-\psi_x)dx,
	\end{multline}
where $X_1$ and $X_2$ are domain wall positions before and after the soliton penetration. Therefore, we have the relation between the soliton profile and the domain wall displacement $\Delta X = X_2-X_1$,
	\begin{equation}
	\Delta X =-\int_{-L/2}^{L/2} (1-\psi_x) dx\approx -\frac{4a}{\lambda+1/2}x_0.
		\label{deltax}
	\end{equation}
This process can occur repeatedly, leading to a discrete manipulation of domain wall. Figure \ref{Fig8}(b) shows the simulation result for a periodically applied stimulation. The stimulation region is from $-500$ nm to $-470$ nm with $A=13$ pJ/m, $\lambda=8$, $w=10$ nm, $d=0$, and $B=4$ T. Outside $-600$ nm and $600$ nm, the damping linearly increasing from 0.0001 to 1 to absorb all kinds of excitations to prevent possible reflection from the boundaries. The stimulation is applied for 1 ps and ceased for 1 ns. A head-to-head domain wall and a tail-to-tail domain wall are initially placed at $x=0$ and $x=300$ nm, respectively. The main figure shows the density plot of $m_x$. The soliton pairs are generated periodically and propagate to both sides with constant speed. After passing through each domain wall, the domain wall shifts towards the stimulation source for a certain distance, regardless of its type. The inset is the close-up of the region indicated by a dashed frame near the domain wall, emphasizing the shift of the domain wall. We then change the stimulation strength $B$ to obtain solitons of different amplitude, and record the displacements $\Delta X$ from the numerical results by finding the position of $m_x=0$ after interpolation. We compare $\Delta X$ with the analytical formula Eq. \eqref{deltax} in Fig. \ref{Fig8}(c), showing good agreement.  
	
The discrete manipulation of domain walls is quite useful in device applications such as the racetrack memory \cite{Parkin2008}, because precise control of domain wall positions is necessary. Our method provides natural discrete domain wall position control and does not need complicated pinning apparatus \cite{pinning} like those continuous-source-driven domain wall motion.  In addition, a strong magnetic field of several Teslas or a current of $10^{13}$ A/m$^2$ lasting 1 ps is not difficult to realize.

\section{Discussion and Summary}
The topologically trivial magnetic solitons have unique properties that are distinct from the well-studied topologically nontrivial magnetic solitons and linear spin waves. Compared to linear spin waves that suffer dispersion, the topologically trivial solitons propagate keeping the localized envelope unchanged. Also, the precessing frequency and the ``wavevnumber" of the trivial soliton are not connected by a dispersion relation like the linear spin waves.
On the other hand, in contrast to nontrivial solitons (such as domain walls and skyrmions) whose structures are nearly rigid,
the trivial ones have internally oscillatory components, so that the Thiele equation \cite{Thiele} is invalid. To see the reaction of the trivial soliton to an external driving, we apply a longitudinal spin-polarized current to a soliton. We consider the adiabatic spin transfer torque \cite{ZhangLi} $\bm{\tau}=v_j \partial \mathbf{m}/\partial x$, where $v_j=Pj\mu_B/eM_s$ ($\mu_B$ is the Bohr magneton, $j$ is the current density, $e$ is the electron charge, and $P$ is the spin polarization). In the presence of this torque (with $\alpha=0$), the soliton velocity changes from $2bv_0$ to $2bv_0+v_j$ \cite{Li2014}. If the current is suddenly turned on, we find that the velocity also changes abruptly, indicating an inertialess dynamics. In this sense, the soliton studied here is massless. A typical result is shown in Appendix C.

The 1D solitons may be experimentally observed in thin magnetic nanostrips. We have numerically verified that the results are qualitatively valid for strip width up to about $2x_0$ ($\sim$10 nm for the parameters we use). For materials with a larger exchange strength, this width can be larger, which is possible to be fabricated \cite{nanowirereview,nanowire}. Imaging of the soliton may be performed using experimental techniques such as X-ray magnetic circular dichroism \cite{RF1} and nitrogen-vacancy center magnetometry \cite{NV1,NV2}. The precession frequency can also be measured by some indirect methods \cite{RF2}. The measured amplitude, speed, and frequency are directly related to $a$ and $b$ in our formula, so that our theoretical results can be experimentally verified.

In summary, the analytical solution of magnetic solitons obtained by using the shape anisotropy approximation, ignoring damping, and assuming small amplitude agrees well with the numerical simulation of the full LLG equation for reasonably low damping and small amplitude. 
When the soliton enters another segment of wire with different anisotropy, it prefers total refraction for weak anisotropy and total reflection for strong anisotropy, with the angular momentum and precession frequency conserved. For intermediate anisotropy near a critical value, the soliton behaves sophisticatedly with the coexistence of refraction and reflection in solitons, as well as conversion to linear spin waves.    
A pair of magnetic solitons can be generated by magnetic field or spin-polarized current pulses of alternating directions applied to at least two successive regions, which is because the magnetic field or spin current of opposite directions creates locally oscillating magnetization profile which is similar to the soliton profile. The generated solitons propagate in opposite directions. Three-region stimulation gives larger amplitude than two-region stimulation when other conditions are the same. The properties of the solitons are controllable by tuning the set-up of the stimulation. A magnetic soliton passing through a domain wall drives the domain wall to shift in the opposite direction of the soliton's propagation, no matter the domain wall is head-to-head or tail-to-tail type. The shift distance can be well described by angular momentum conservation. These findings enable a proposal for the discrete manipulation of domain walls, and provide a paradigm for magnetic soliton-based devices.

\begin{acknowledgements}
This work was supported by the National Natural Science Foundation of China (Grants No. 11804045 and 12174093), the Natural Science Foundation of Hunan Province of China (Grant No. 2025JJ60001), and the Fundamental Research Funds for the Central Universities.
\end{acknowledgements}

\section*{Appendix A: A soliton converting to a refracted soliton and a reflected linear spin wave packet}

\begin{figure}[!ht]
		\centering
		\includegraphics[width=\textwidth]{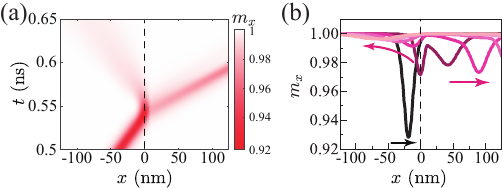}
		\caption{(a) Density plot of $m_x$ in $x-t$ plane near the interface for $\lambda_1=7$, $\lambda_2=5.5$, and incident soliton $a=0.5$, $b=0.5$. (b) Snapshots of $m_x$ for the incident soliton (black line) and the profiles after passing the interface (pink lines with different brightness). The higher brightness corresponds to later time.}
		\label{append}
	\end{figure}

Here we show a typical example that a soliton refracts in the form of soliton, but reflects in the form of linear spin wave packet. The case of $\lambda_1=7$, $\lambda_2=5.5$, and incident soliton $a=0.5$, $b=0.5$ is shown in Fig. \ref{append}. From the density plot of $m_x$ in (a) and the snapshots at different time in (b), it is clear that the refracted part is still a soliton, but the reflected part has obvious dispersion, meaning it is a linear spin wave packet. The conversion from soliton to linear spin wave is actually ubiquitous in the numerical studies in Sec. \ref{interface}. But for relatively weak and strong $\lambda_2$ far from the critical value, the dissipation into linear spin wave is negligible.

    \section*{Appendix B: Reflection and refraction of a linear spin wave in 1D}
For a linear spin wave, the nonlinear term in Eq. \eqref{NLS} is further ignored. Thus, the reflection and refraction of 1D spin wave at an interface of different anisotropies are the same as the electron hitting a step potential. The dispersion relation of linear spin waves is
\begin{equation}
    \omega=\frac{2\gamma}{M_s}\left(Ak^2+K\right).
\end{equation}
We assume that the incident spin wave is $c_0 e^{i(kx-\omega t)}$, the reflected wave is $c_1e^{i(-kx-\omega t)}$, and the refracted wave is $c_2e^{i(k'x-\omega t)}$, then the continuity conditions ($\psi$ and $\partial_x \psi$ are continuous) give $c_0+c_1=c_2$ and $kc_0-kc_1=k'c_2$.
Thus, the amplitudes of reflected and refracted spin waves are
\begin{equation}
c_1=\frac{k'-k}{k'+k}c_0,\quad c_2=\frac{2k}{k'+k}c_0.
\end{equation}
For the $a=0.3$, $b=1$ soliton we discuss in Fig. \ref{Fig3}, the precessing frequency is $\omega_0=(\gamma \mu_0M_s)[\lambda_1-(a^2-b^2)]$. For a linear spin wave with the same frequency, the incident wavevector $k$ is $0.188x_0^{-1}$. For $\lambda_2=6$, $k'=0.272 x_0^{-1}$, so the reflected amplitude $c_1=-0.183$. This is different from the invisible reflection in Fig. \ref{Fig3}(a). Also, the threshold $\lambda_2$ for total reflection is $\lambda_2>\lambda_1-(a^2-b^2)=7.91$ ($k'$ becomes imaginary), which is also different from the value 8.02 we have got for solitons. This can be seen from the data point of $\lambda_2=8$ in Fig. \ref{Fig3}(d), where obvious refraction can still be observed.
Of course, because the linear spin waves are extended rather than localized, the reflection and refraction cannot be visualized in the form of density plots in Fig. \ref{Fig3}.

 \section*{Appendix C: Dynamics of a soliton under adiabatic spin transfer torque}
 \begin{figure}[!ht]
		\centering
		\includegraphics[width=\textwidth]{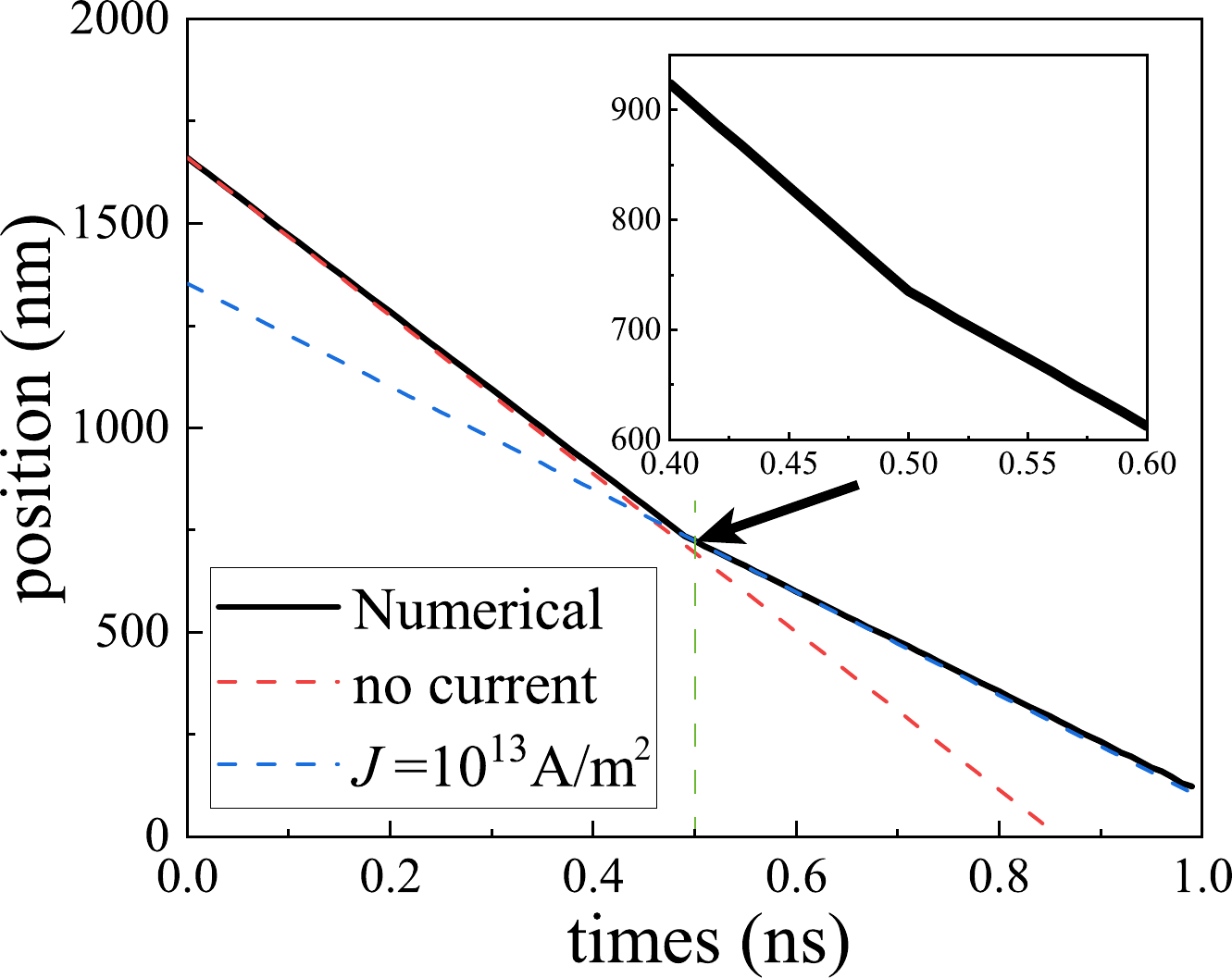}
		\caption{Motion of a soliton of $a=0.3$, $b=1$ driven by adiabatic spin transfer torque turned on at 0.5 ns with $\alpha=0$. The solid line is the numerical result, and the dashed lines are results of analytical formula. Inset: Close-up near 0.5 ns.}
		\label{STTdriven}
	\end{figure}	
Here, we use the same material parameters as those in Fig. \ref{Fig2}(b), and $\alpha$ is set to be 0. We use an $a=0.3$, $b=1$ soliton as the initial state, and turn on a fully polarized current of $j=10^{13}$ A/m$^2$ ($v_j=674.0$ m/s) at 0.5 ns. The position of the soliton is shown in Fig. \ref{STTdriven}. The inset is a close-up near 0.5 ns with time step chosen to be less than 1 ps. The sharp turning of the slope indicates that the velocity changes abruptly without any acceleration process. Further investigations about the driven dynamics of the topologically trivial solitons are necessary.
    \bibliographystyle{iopart-num.bst} 
	\bibliography{magneticsoliton.bib}
\end{document}